\documentclass[prl,twocolumn,superscriptaddress,showpacs,floatfix]{revtex4-2}

\usepackage[utf8]{inputenc}
\usepackage{graphicx}
\usepackage{dcolumn}
\usepackage{braket}
\usepackage{bm}
\usepackage{amsfonts} 
\usepackage{amsmath} 
\usepackage[bookmarksnumbered,pdfpagelabels=true,plainpages=false,colorlinks=true,linkcolor=blue,citecolor=blue,urlcolor=blue]{hyperref}
\usepackage[bookmarksnumbered,pdfpagelabels=true,plainpages=false,colorlinks=true,linkcolor=blue,citecolor=blue,urlcolor=blue]{hyperref}

\def \usach {Departamento de F\'isica, Universidad de Santiago de Chile, 9170124, Santiago, Chile.}
\def \cedenna {Centro  de Nanociencia y Nanotecnología CEDENNA, Avda. Ecuador 3493, Santiago, Chile.}
\def \fcfm {Departamento de F{\'i}sica, CEDENNA; Facultad de Ciencias Físicas y Matemáticas,  Universidad de Chile, Santiago, Chile.}

\def \cbr {Hitachi Cambridge Laboratory, J. J. Thomson Avenue, Cambridge CB3 0HE, United Kingdom}

\begin{document}

 \title{Moving Analogue Horizons in Stationary Ferroelectrics}

\author{David Galvez-Poblete}
\email{david.galvez.p@usach.cl}
\affiliation{\usach}
\affiliation{\cedenna}

\author{Rubén M. Otxoa}
\affiliation{\cbr}

\author{Alvaro S. Nunez}
\affiliation{\fcfm}

\author{Sebastian Allende}
\affiliation{\usach}
\affiliation{\cedenna}

\begin{abstract}

We show that a traveling modulation of the polarization-gradient stiffness in a ferroelectric material induces an effective flow in its collective polarization dynamics. Within a controlled local approximation, small polarization fluctuations, or ferrons, obey a massive Klein-Gordon equation with flow. Unlike conventional analogue-gravity platforms, the effective flow originates from the modulation of the material parameters rather than from the physical transport of the medium. This mechanism enables mobile analogue horizons separating sub-ferronic and super-ferronic regions, the latter supporting negative-norm antiferron modes. Their coupling to positive-norm ferrons gives rise to superradiant-like amplification, while the effective mass gap can be tuned independently through an external electric field. Ferroelectric systems therefore provide a novel and experimentally controllable platform for analogue gravity with massive scalar excitations. 

\end{abstract}

\maketitle

\section*{Introduction}

Ferroelectric materials inherently possess broken inversion symmetry\cite{Anderson1997} and exhibit strongly anharmonic energy landscapes\cite{Rabe2007}, supporting a wide variety of collective polarization dynamics. In recent years, these systems have emerged as highly tunable condensed-matter platforms, in which the polarization state and its collective excitations can be manipulated by external electric fields, strain engineering, interfaces, heterostructures, and ultrafast driving protocols. Such versatility makes ferroelectric systems particularly attractive for studying nonequilibrium and wave-propagation phenomena in controllable environments\cite{Tang2022, RodriguezSuarez2024, Pols2026, Choe2026}.

Small fluctuations of the polarization field around the ferroelectric equilibrium configuration, which we refer to as ferrons, provide a natural framework to study collective excitations in these systems. As we show in this work, within the Landau–Ginzburg–Devonshire (LGD) formalism\cite{Chandra2007, Levanyuk2020, Li2005, Morozovska2011, Cao2008, Castro2025}, the dynamics of ferrons can be mapped onto a massive Klein–Gordon equation, establishing a direct connection between ferroelectric collective modes and relativistic scalar-field dynamics.

The notion that wave propagation in moving media can mimic the kinematics of fields on curved spacetimes dates back to the seminal proposal of Unruh~\cite{Unruh1981}, who showed that sound waves in a transonic fluid flow obey a wave equation formally identical to that of a massless scalar field on a Lorentzian black-hole geometry, with the sonic horizon playing the role of the event horizon and giving rise to a thermal flux of phonons analogous to Hawking radiation. This acoustic metric was placed on rigorous footing by Visser~\cite{Visser1998} and generalized into a broader analog-gravity program encompassing electromagnetic, gravitational, and quantum-fluid media~\cite{Visser2002Models}, comprehensively surveyed in the Living Reviews article by Barcel\'o, Liberati and Visser~\cite{Barcelo2011} and in the monographs of Novello, Visser and Volovik~\cite{NovelloVisserVolovik2002} and Volovik~\cite{Volovik2009}. Bose-Einstein condensates emerged as a particularly versatile experimental platform, following the theoretical proposals of Garay \emph{et al.}~\cite{Garay2000} and Barcel\'o, Liberati and Visser~\cite{Barcelo2003} and the numerical study of Carusotto \emph{et al.}~\cite{Carusotto2008}; sonic horizons were realized experimentally by Lahav \emph{et al.}~\cite{Lahav2010}, leading to the observation of black-hole-laser self-amplification~\cite{Steinhauer2014} and, subsequently, of spontaneous Hawking radiation together with its entanglement signature~\cite{Steinhauer2016}, a result whose statistical interpretation was later debated~\cite{Leonhardt2018, Steinhauer2018Comment} and refined through measurements of the Hawking temperature~\cite{MunozdeNova2019} and of the stationary, time-resolved emission spectrum~\cite{Kolobov2021}. Parallel realizations were achieved in nonlinear optical fibers, where a co-propagating soliton creates an effective event horizon for probe light~\cite{Philbin2008}, and in classical surface-wave hydrodynamics, where stimulated and spontaneous Hawking-like scattering has been measured in water-tank flows~\cite{Weinfurtner2011, Euve2016}. Superfluid $^3$He-A offers a further low-temperature realization, with moving domain walls and vortex textures generating effective horizons for Bogoliubov quasiparticles~\cite{JacobsonVolovik1998, KopninVolovik1998}. More recently, the analog-gravity framework has been extended to magnetic systems, where Roldán-Molina, Núñez, and Duine showed that magnon dynamics on inhomogeneous, precessing spin textures can be mapped onto wave propagation on an effective curved spacetime with an associated magnonic event horizon~\cite{RoldanMolina2017, Doornenbal2019}. These ideas, along with the notion of a magnonic white hole, led to the intriguing concept of a Black hole-White hole topological magnonic crystal\cite{GalvezPoblete2026}. The role of vorticity, largely absent from the original irrotational-flow constructions, has been examined systematically by Cropp, Liberati, and Turcati~\cite{Cropp2016} and by Liberati, Schuster, Tricella, and Visser~\cite{Liberati2019Vorticity}, clarifying the conditions under which rotating astrophysical spacetimes can be faithfully reproduced in the laboratory. For a forward-looking assessment of the field and a historical literature review, see Jacquet, Weinfurtner and K\"onig~\cite{Jacquet2020} and Almeida and Jacquet~\cite{Almeida2023}.


A central ingredient in many analog-gravity systems is the presence of an effective flow, which induces a Doppler-like shift\cite{Lee2010, Vlaminck2008, FernandezRossier2004, Zawislak2025, Reed2003, Ginzburg1996} in the excitation spectrum and can lead to negative-energy excitations. In conventional platforms, this effect is typically associated with the transport of a certain physical quantity.

In this work, we develop an effective theoretical description of a ferroelectric system in which the effective flow arises from traveling modulations of the material parameters themselves. By considering dynamical spatial modulations of the Landau-Ginzburg-Devonshire stiffness and Landau parameters, we show that ferronic excitations experience an effective convective dynamics, leading to a Doppler-shifted dispersion relation. Furthermore, we discuss the validity of the local approximation underlying the emergent-flow picture, and analyze the conditions for the appearance of analogue horizons and superradiance-like phenomena in ferroelectric platforms.

\section*{Generalized Landau-Ginzburg-Devonshire Framework}

To describe the dynamics of ferroelectric materials, we adopt the generalized Landau-Ginzburg-Devonshire (LGD) framework in one spatial dimension, including spatial variations of the polarization field \( P(x,t) \). The free energy functional is \cite{Cao2008,Bain2017}:
\begin{equation}
F[P(x)] = \int dx \, \mathcal{F}(P, \partial_x P, t),
\end{equation}
where the free energy density \(\mathcal{F}\) is given by:
\begin{equation}
\mathcal{F} = \frac{a}{2} P^2 + \frac{b}{4} P^4 + \frac{c}{6}P^6 + \frac{D}{2}\left(\frac{\partial P}{\partial x}\right)^2 - E(t) P.
\end{equation}

Here, the gradient term \(\frac{D}{2}(\partial_x P)^2\) penalizes sharp spatial variations, representing domain-wall energy and enabling spatially coherent collective modes. For simplicity, in this work we will consider $c = E(t) =0$, but the results are extensible for other choices of these parameters.

We introduce the Lagrangian density for the polarization field:
\begin{equation}
\mathcal{L}(P, \dot{P}, t) = \frac{\rho}{2} \dot{P}^2 - \mathcal{F}(P, \partial_x P, t),
\end{equation}

where \(\rho\) is the effective inertia parameter per unit length. The equation of motion for the system is given by the Landau-Khalatnikov-Tani equation \cite{Ishibashi1989,Widom2010}:
\begin{equation}
    \rho \frac{\partial^2}{\partial t^2}P = - \frac{\delta \mathcal{F}}{\delta P}
\end{equation}

The stable uniform configuration is obtained by minimizing the free energy density, which yields $P_{o} = \pm  \sqrt{-a/b}$ for $a<0$. We consider small fluctuations around the equilibrium configuration of the form $ P(x,t) = P_o + \phi (x,t)$, with $|\phi| \ll P_o$. Therefore, the equation of motion of these fluctuations, which we refer to as ferrons, is given by:
\begin{equation}
   \partial_t^2 \phi - v^2 \partial_x ^2 \phi  + m^2 \phi =0 
\end{equation}

Here, we have neglected terms of order $\phi^2$ and defined the velocity as $v^2 \equiv D/\rho $ and the effective mass as $m^2 \equiv (a + 3bP_o^2)/\rho$, which determines the gap of the acoustic ferroelectric modes, $\omega_{gap} = m$ . This equation also allows us to define a characteristic length scale, $L_c \equiv v/m$. The resulting equation corresponds to a massive scalar Klein-Gordon, which is commonly associated with relativistic quantum fields describing spinless particles. Therefore, this system provides a natural framework for studying analogue gravity.

We now consider a dynamical stiffness parameter, $D(x,t)$. This modifies the equation of motion of the ferroelectric system as follows:
\begin{equation}
\rho \partial_t^2 P - \partial_x \left[ D(x,t) \partial_xP \right] + aP + bP^3 =0
\end{equation}

If the stiffness and rigidity vary as traveling modulation, $D(x,t) = D_o + \delta D f (Kx-\Omega t)$ and $a(x,t ) = a (Kx-\Omega t)$, we can define a co-moving coordinate, $\xi \equiv x-t \Omega/K$, and the velocity of the modulation $u \equiv \Omega/K$. It is important to note that $a<0$ at all times to guarantee the stability of the system. Under these considerations, a good approximation for the stationary configuration is given by:
\begin{equation}
    P_o(\xi) = \sqrt{-a(\xi)/b} 
\end{equation}

The stationary configuration $P_0(\xi) = \sqrt{-a(\xi)/b}$ corresponds to the local minimum of the Landau free energy density evaluated at each point $\xi$. To assess its validity as an approximate solution of the full equation of motion, we substitute it into the static limit of Eq.~6 in the co-moving frame. The Landau terms $a(\xi)P_0 + bP_0^3$ cancel exactly by construction, leaving a residual gradient contribution of order $D/(L^2)\cdot P_0$, where $L$ is the characteristic length scale over which $a(\xi)$ and $D(\xi)$ vary. 
The corrected background takes the form $P_0(\xi) \to P_0(\xi) + \delta P(\xi)$, where
\begin{equation}
    \delta P(\xi) \sim \frac{D}{L^2|a|}P_0 \equiv \epsilon^2 P_0, 
    \qquad \epsilon \equiv KL_c = \sqrt{\frac{D}{|a|L^2}},
\end{equation}
so that $|\delta P|\ll P_0$ requires the adiabaticity condition $\epsilon\ll 1$, i.e., the modulation wavelength $L = 2\pi/K$ must be much larger than the 
characteristic ferronic length $L_c = \sqrt{D/|a|}$. Introducing fluctuations via $P(x,t) = P_0(\xi) + \delta P(\xi) + \phi(\xi,t)$ 
and linearizing in $\phi$, the equation of motion becomes

\begin{equation}
    \rho(\partial_t - u\partial_\xi)^2\phi 
    - \partial_\xi\left[D(\xi)\partial_\xi\phi\right] 
    + \rho m^2(\xi)\,\phi = \mathcal{O}(\epsilon^2)\,\phi,
    \label{eq:ferron_full}
\end{equation}
where $m^2(\xi) = -2a(\xi)/\rho$ is the local effective mass squared. 
The correction arising from the background perturbation $\delta P$ enters at 
relative order $3\epsilon^2$, consistently with the adiabatic expansion. 
Restricting to a region where $D(\xi)$, $a(\xi)$, and $P_0(\xi)$ vary slowly 
on the scale of $\lambda_\phi$, the coefficients may be treated as locally 
constant and Eq.~(\ref{eq:ferron_full}) reduces to the massive Klein-Gordon 
equation with flux
\begin{equation}
    (\partial_t - u\partial_\xi)^2\phi 
    - v^2\partial_\xi^2\phi 
    + m^2\phi = 0,
    \label{eq:KG_flux}
\end{equation}
with $v^2 = D/\rho$ and $m^2 = -2a/\rho$ evaluated locally. 
The two conditions, validity of the background configuration $P_0(\xi)$ 
and validity of the local metric description, are therefore not independent: 
both are controlled by the single adiabaticity parameter $\epsilon$, 
with corrections entering uniformly at order $\epsilon^2$. 
The framework is thus internally self-consistent within the regime $\epsilon\ll 1$, 
which constitutes a single self-consistency requirement on the modulation protocol.
This equation describes locally the dynamics of fluctuation of the polarization with wavelengths much lower than the characteristic length of the system $\lambda_\phi \ll L_c \ll 2\pi/K$. The $u \partial_\xi$ term induces a Doppler shift in the dispersion relation, which is crucial for defining an analogue horizon. In this case, the effective flow is induced through a dynamical modulation of the system parameters, acting as an effective moving medium through which the excitations can propagate. The dynamical modulation of the gradient coefficient $D$ is physically realized through the flexoelectric coupling mechanism, which renormalizes the polarization gradient coefficient\cite{Morozovska2016,Eliseev2009, Morozovska2017}. A traveling surface acoustic wave provides a spatiotemporal strain field that extends this renormalization to the dynamical case, naturally implementing the traveling modulation required for the analogue horizon.

\begin{figure}
    \centering
    \includegraphics[width=1.0\linewidth, trim = 6cm 7cm 7cm 6cm]{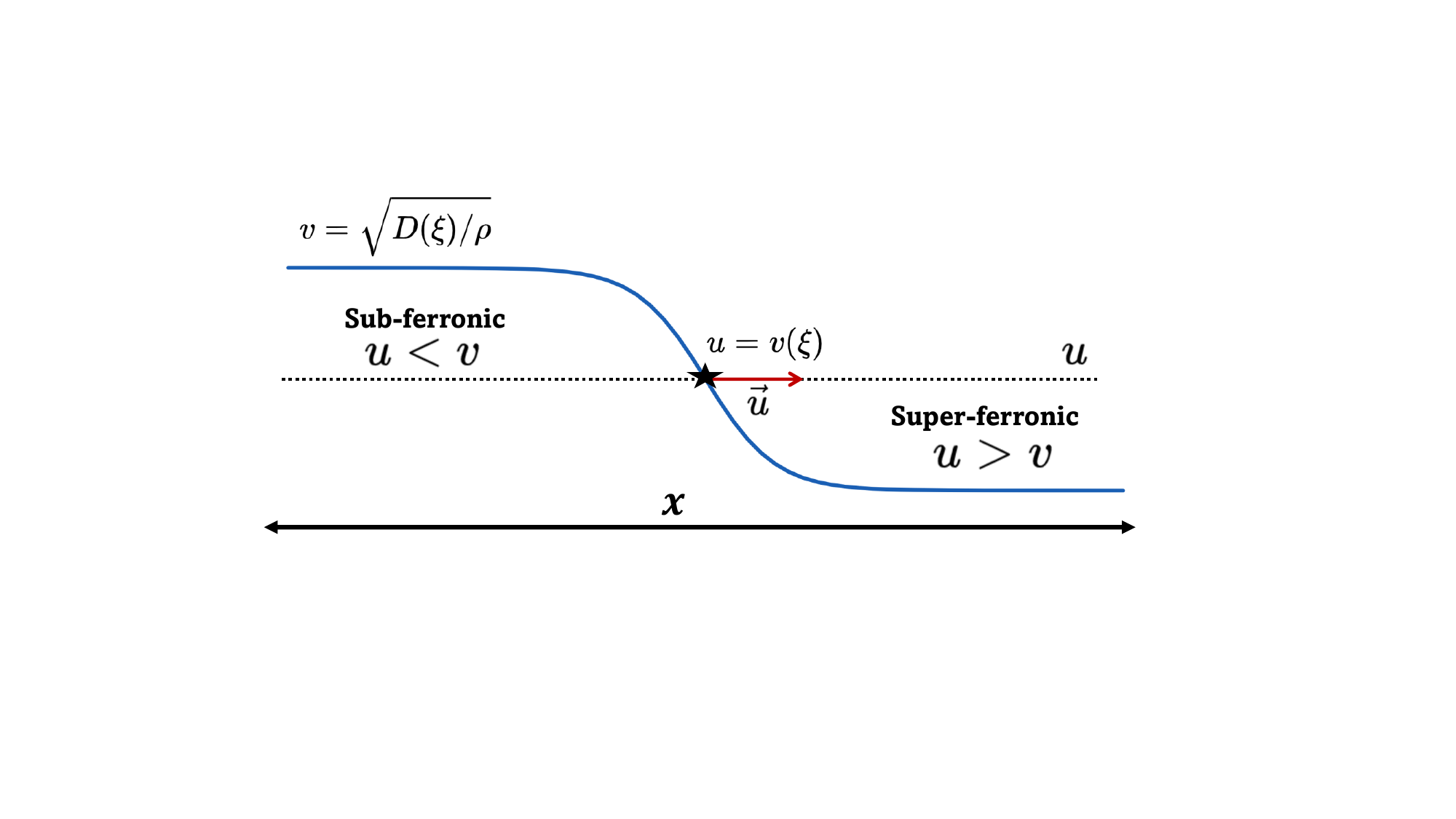}
    \caption{Schematic of dynamic modulation of the ferronic velocity $v = \sqrt{D(\xi)/\rho}$. The analogue horizon, where $u = v (\xi)$, is represented by the star, which moves with velocity $u$ in the laboratory frame. Here $f(K\xi)$ is taken as a hyperbolic tangent profile.}
    \label{fig:placeholder}
\end{figure}

\section{Analogue horizon}
Using Eq. 10 as a starting point, we can derive the conditions required for the emergence of analogue horizons, depending on the relation between $u$ and $v$. To analyze this, we first determine the dispersion relation of the small excitations of the system. By proposing plane-wave solutions, we obtain:
\begin{equation}
    (\omega -uk)^2 = v^2k^2 + m^2
\end{equation}

\begin{figure*}[t]
    \centering
    \includegraphics[width=\textwidth]{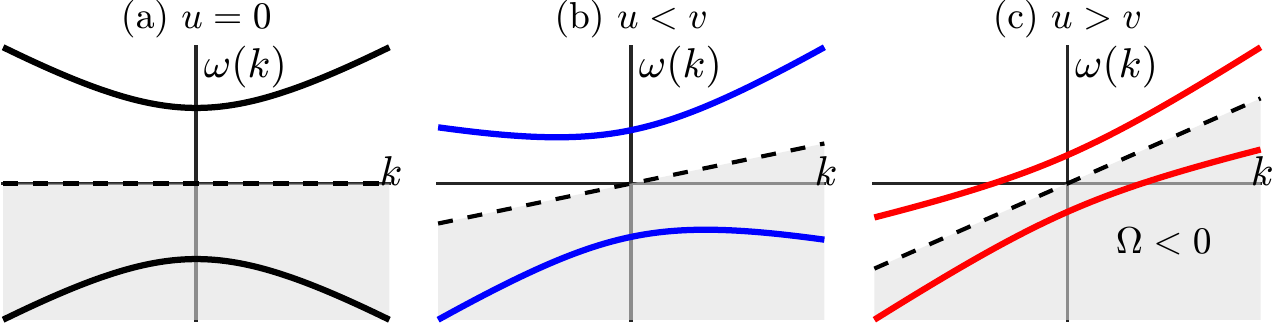}
    \caption{Dispersion relation of ferronic excitations. (a) Zero-flow case, showing the gapped ferronic and antiferronic branches. (b) Subferronic regime ($u<v$), where the Doppler shift modifies the dispersion while preserving positive norm modes. (c) Superferronic regime ($u>v$), where a portion of the spectrum enters the $\Omega=\omega-uk<0$ region, giving rise to negative-norm modes. The dashed line corresponds to $\Omega=0$.}
    \label{fig:dispersion}
\end{figure*}

The massive KG equation with flow possesses a conserved current. For a plane-wave solution, this conserved current reduces to:
\begin{equation}
    \mathcal{J} = 2 |A|^2 \left( \omega u - (u^2 - v^2 ) k \right)
\end{equation}

And the norm of the mode is determined by the sign of the co-moving frequency:
\begin{equation*}
    sign(\Omega) = sign(\omega- uk)
\end{equation*}

Therefore, modes with $\Omega <0$ carry negative norm and correspond to antiferron excitations \cite{GalvezPoblete2025}.

Neglecting the mass term locally, Eq. 10 can be written in the form of a Klein-Gordon equation in curved spacetime,

\begin{equation}
\frac{1}{\sqrt{-g}}
\partial_\mu
\left(
\sqrt{-g}\,
g^{\mu\nu}
\partial_\nu \phi
\right)
=0.
\end{equation}

By comparing both expressions, one identifies the effective inverse metric tensor, up to an irrelevant conformal factor, as
\begin{equation}
g^{\mu\nu}
\propto
\begin{pmatrix}
1 & u\\
u & u^2-v^2
\end{pmatrix}.
\end{equation}

The corresponding covariant metric tensor is therefore

\begin{equation}
g_{\mu\nu}
\propto
\begin{pmatrix}
-(v^2-u^2) & u\\
u & -1
\end{pmatrix},
\end{equation}

which yields the effective line element $ds^2=-(v^2-u^2)dt^2+2u\,dt\,dx-dx^2.$ This metric corresponds to a Painlevé-Gullstrand-like geometry \cite{Visser1998}, where the analogue horizon appears when $g_{tt}=0$, which implies $u=v$.

When $u<v$, the system supports both propagating and counter propagating modes, although an intrinsic asymmetry between them is induced by the flow. In this regime, both modes are associated with positive-energy (positive-norm) solutions, as can be seen from the upper branch of the blue curve in Fig 2, for both $k<0$ and $k>0$. In contrast, when $u>v$, the system no longer allows counter-propagating modes and instead exhibits two co-propagating modes, one of which is associated with negative-energy (negative-norm) solutions, which we identify as antiferron modes. These correspond to the two positive branches of the red curve in Fig.2.    

At the transition between the two region, where $u =v$, i.e, when the modulation propagates at the same velocity as ferrons in the material, we can define an analogue ferroelectric horizon. It is important to note that the condition $u=v(\xi)$, which defines the analogue horizon, is satisfied in the co-moving frame, the stiffness profile $D(\xi)$ propagates at velocity $u$, and therefore the horizon moves together with the modulation as we can observe in fig.1. As a result, the system supports a traveling analogue horizon.

If the effective flow is directed along the $\hat{x}$ direction, and the left region satisfies $u<v$, which we define as the subferronic region, while the right region satisfies $u>v$, which we define as the superferronic region, then interface defined by $u=v$ corresponds to an analogue ferroelectric black hole horizon. This occurs because propagating excitation modes cannot escape from the superferronic region into the subferronic one.

Conversely, if the superferronic region is located on the left and the subferronic region on the right, the interface between them, defined by $u=v$, corresponds to an analogue ferroelectric white-hole horizon. This means that a ferronic excitation originating in the subferronic region cannot enter the superferronic one.

It has been previously shown that in this kind of the system, where we have finite length cavities bounded by analogue horizons, the scattering properties may exhibits superradiance-like phenomena. For example, in an WH-BH pair configuration, the two transmitted mode satisfies \cite{GalvezPoblete2026}:
\begin{equation}
    \mathcal{T}_{+} - \mathcal{T}_{-} = 1
\end{equation}
Where the hybridization of positive and negative energy modes can give transmission coefficients greater than unity.

From an experimental perspective, the most favorable strategy is not necessarily to increase the modulation velocity \(u=\Omega/K\), but to engineer regions with reduced ferron velocity \(v=\sqrt{D/\rho}\). A traveling modulation with experimentally accessible phase velocity may then become super-ferronic in the softened region, producing an analogue horizon whenever \(u=v(x_h)\). This can be pursued either through smooth stiffness gradients, where the local metric description applies, or through abrupt heterostructure interfaces, where the problem is naturally formulated as a scattering problem between homogeneous regions.

For a smooth analogue horizon, one can estimate the associated Hawking temperature from the analogue surface gravity. In particular, for a black-hole horizon, the temperature is given by \cite{2013}
\begin{equation}
    T_H = \frac{\hbar}{2\pi k_B}\kappa, 
    \qquad \kappa = \left|\partial_\xi(u - v)\right|_{\xi_h},
\end{equation}
where $\kappa$ is the analogue surface gravity evaluated at the 
horizon position $\xi_h$. Since the modulation velocity $u$ is 
spatially uniform and the horizon is produced by a smooth spatial 
variation of the ferronic velocity $v(\xi) = \sqrt{D(\xi)/\rho}$, 
the surface gravity reduces to $\kappa \simeq |dv/d\xi|_{\xi_h}$. 
Assuming that $v(\xi)$ varies from $v_0$ to $v_\mathrm{min} = 
v_0\sqrt{1 - \delta D/D_0}$ over a characteristic horizon length 
$L$, one obtains
\begin{equation}
    T_H \simeq \frac{\hbar}{2\pi k_B}
    \frac{v_0\left(1 - \sqrt{1-\delta D/D_0}\right)}{L}.
    \label{eq:TH}
\end{equation}

A critical consistency condition must be imposed: the Landau 
coefficient $a(\xi)$ must remain strictly negative throughout 
the entire spatial profile, including at the horizon. Any spatial 
variation with $\delta a/|a_0| \gtrsim 1$ would drive the system 
across a local ferroelectric phase transition, whose critical 
fluctuations would completely mask the analogue Hawking signal. 
This strongly favors experimental configurations where the horizon 
is produced exclusively by modulating $D(\xi)$, while keeping $a$ 
approximately uniform. Under this condition, the effective mass 
$\tilde{m}^2 = -2a/\rho$ remains finite and uniform across the 
horizon, ensuring the validity of Eq.~(\ref{eq:TH}).

To assess experimental feasibility, we analyze the parameter 
requirements for LiNbO$_3$, using the LGD parameters \cite{Scrymgeour2005}. The ferronic velocity is $v_0 = \sqrt{D_0/\rho} 
\approx 17.3$ km/s and the characteristic ferronic length is 
$L_c = \sqrt{D_0/(-2\alpha)} \approx 0.37$ nm. Since the Curie 
temperature of LiNbO$_3$ is $T_c \approx 1463$ K, both room 
temperature and any cryogenic operating temperature satisfy 
$T \ll T_c$, placing the system deep in the ferroelectric phase 
in both cases. The LGD parameters therefore vary negligibly 
between room temperature and the millikelvin regime, and the 
tabulated room-temperature values provide an accurate description 
of the system at the operating temperature $T_\mathrm{bath} 
\sim 0.1$ K.

The horizon is engineered by a spatial profile $D(\xi)$ that 
decreases smoothly from $D_0$ to $D_0(1 - \delta D/D_0)$ over 
a length $L$, so that the local ferronic velocity matches the 
modulation velocity $u $ at the horizon. The 
modulation is implemented as a traveling wave with wavelength 
$\lambda_\mathrm{mod}$, giving a required frequency 
$\Omega/2\pi = u/\lambda_\mathrm{mod}$. For a choice of 
$\lambda_\mathrm{mod} = 1~\mu$m, the adiabaticity parameter 
is $\epsilon = KL_c \ll 1$, confirming that the local 
approximation is excellently satisfied. The required frequency 
falls in the GHz range, within the operational window of 
LiNbO$_3$-based surface acoustic wave resonators \cite{Huang2023}.

Figure 3 shows $T_H$ as a function of $L/L_c$ for four values 
of $\delta D/D_0$, over the range $5 \lesssim L/L_c \lesssim 50$ 
where the continuum LGD description is valid.
Over this range, $T_H$ remains 
well above the bath temperature $T_\mathrm{bath} \sim 0.1$ K 
achievable in standard dilution refrigerators, with 
$T_H/T_\mathrm{bath} \gg 1$ throughout.


For the Hawking temperatures estimated here, the characteristic frequency scale $\omega_H = k_BT_H/\hbar$ falls in the GHz range. An important consequence of the massive ferroic gap, $m_\text{eff}^2 = - 2 a /\rho$, is the suppression of Hawking radiation at the analogue horizon. In the sub-ferronic region, the frequency gap, $\omega_{gap}^{sub} = m \sqrt{1-u^2/v^2}$, is much larger than the characteristic Hawking frequency $\omega_H$. Consequently, the Hawking radiation is exponentially suppressed in this region. On the other hand, although the superferronic region is gapless due to the effective flow and, in principle, the Hawking radiation could propagate through both available channels (positive and negative norm modes), the Doppler transformation of the radiation spectrum induced by the motion of the horizon leads to an additional dynamical suppression of Hawking radiation.  Therefore, the observable signature of this system is not expected to be a thermal Hawking spectrum. Instead, the most relevant signatures are associated with superradiant-like amplification processes arising from the coupling between positive and negative norm modes.


An important feature of this system is that the effective mass gap is fully tunable through and external electric field. The application of an electric field modifies the equilibrium polarization $P_o(\xi)$:
\begin{equation*}
    a P_o + bP_o^3 = E
\end{equation*}
and, consequently, changes the effective mass that determines the gap: 
\begin{equation*}
    m_{\text{eff}}^2 = (a+3bP_o^2)/\rho  
\end{equation*}
 Therefore, this system represents a novel platform to studying a mobile analogue horizon with a tunable mass controlled by an external electric field.

Finally, it is important to emphasize that the formalism presented in this work is applicable to any modulation $f(K \xi)$ of the gradient coefficient. Thus, a dynamical modulation of the form $f (Kx-\Omega t) = \cos (Kx-\Omega t)$ would generate a periodic array of traveling analogue horizons. Such a configuration could provide a platform for investigating the dynamics of analogue black-hole-white-hole crystals  and associated superradiance band structures.

\begin{figure}
    \centering
    \includegraphics[width=0.5\linewidth, trim = 7cm 9cm 7cm 10cm]{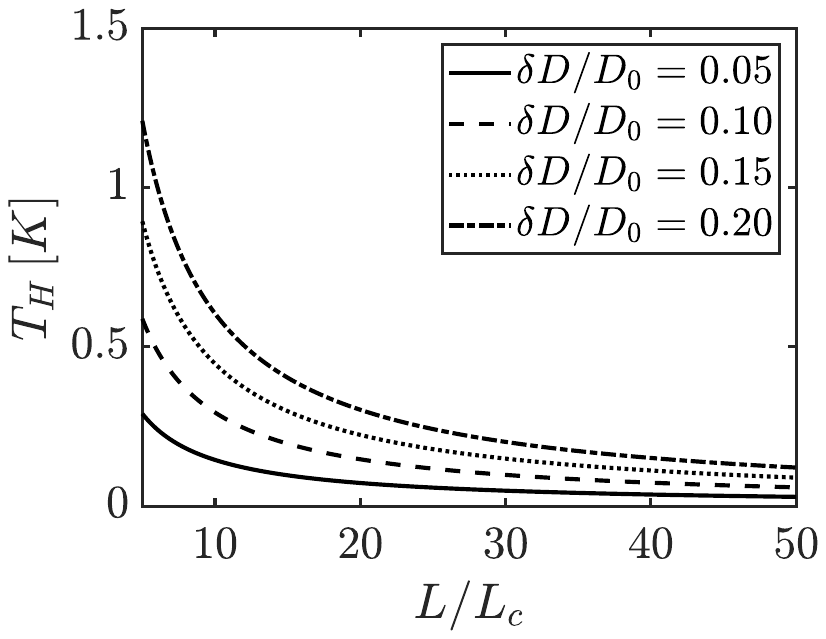}
    \caption{Hawking temperature of the ferroelectric $\text{LiNbO}_3$ analogue horizon as a function of the dimensionless transition length $L/Lc$ for different modulation amplitudes.}
    \label{fig:placeholder}
\end{figure}

\section*{Conclusions}
In this work, we study a one-dimensional ferroelectric system within the Landau-Ginzburg-Devonshire(LGD) formalism, using the Landau-Khalatnikov-Tani equation to describe the polarization dynamics. By analyzing small fluctuations around the ferroelectric equilibrium state, we show that the effective dynamics of the polarization excitations can be described by a massive Klein-Gordon equation.

Introducing a traveling modulation of the stiffness and rigidity, and within the local approximation, we obtain an effective Klein-Gordon dynamics with an induced flow term. Depending on the relation between the flow velocity and the local ferronic velocity, the system may support negative-norm ferronic modes.

Furthermore, by considering spatial variations of the ferronic velocity, which may be engineered through experimentally accessible setups, we derive the condition for the formation of analogue horizons in ferroelectric systems. Such configurations are of particular interest because they may exhibit superradiance-like phenomena, as illustrated here for a bounded cavity configuration. We also determine the Hawking temperature for typical ferroelectric material, $\text{LiNbO}_3$, obtaining values in the range of $0.1-1.0 \text{K}$. 

Finally, throughout this work we have set $E = 0$ for simplicity. The inclusion of an external electric field introduces qualitatively new physics that deserves separate consideration. 
A static but spatially varying field $E(x)$ modifies the equilibrium polarization $P_0$ and therefore the effective mass profile independently of the stiffness modulation $D(\xi)$, providing an additional and decoupled control 
parameter over the system. Furthermore, a time-periodic uniform field $E(t) = E_0\cos(\omega_d t)$ near the parametric resonance condition $\omega_d \approx 2\omega_\mathrm{gap}$ 
induces a Mathieu-like instability in the ferronic fluctuations, which could selectively amplify modes near the horizon and potentially 
enhance the detectability. These possibilities are left for future work.

Ferroelectric systems thus constitute a novel condensed-matter platform for mobile analogue horizons in massive scalar fields, with tunable mass gap controlled by an external electric field and superradiant-like amplification as the primary observable signature.

\section*{Acknowledgement}
S.A. acknowledges funding from Fondecyt Regular 1261323 and ANID Cedenna CIA250002. A.S.N. acknowledges funding from Fondecyt Regular 1230515 and ANID CEDENNA CIA250002.  D. G.-P. acknowledges ANID-Subdirección de Capital Humano/Doctorado Nacional/2023-21230818.

\bibliography{ferroBH}

\end{document}